\documentclass{elsart}

\usepackage{graphicx}
\usepackage{amssymb}

\def\ltsim{\footnotesize{\mathop{\raisebox{-.4ex}{\rlap{$\sim$}}
\raisebox{.4ex}{$<$}}}}

\renewcommand{\theta}{\vartheta}
\renewcommand{\epsilon}{\varepsilon}

\begin{document}

\begin{frontmatter}


% \thanks[label2]{}http://nemoweb.lns.infn.it/

% \title{Possible detection of TeV neutrinos from Galactic Microquasars with an
% underwater \v{C}erenkov km${\bf ^3}$ telescope}

\title{Sensitivity of an underwater \v{C}erenkov km${\bf ^3}$ telescope
to TeV neutrinos from Galactic Microquasars}

%%% Autor List: Name Orderiring

 \author[INFNCT]{S. Aiello},
 \author[INFNBA]{M. Ambriola},
 \author[INFNRM]{F. Ameli},
 \author[INFNLNS,UniCT]{I. Amore},
 \author[INFNGE]{M. Anghinolfi},
 \author[INFNLNS]{A. Anzalone},
% \author[INFNPI]{C. Avanzini},
 \author[INFNNA]{G. Barbarino},
 \author[INFNBA]{E. Barbarito},
 \author[INFNGE]{M. Battaglieri},
 \author[INFNBA]{R. Bellotti},
 \author[INFNPI]{N. Beverini},
 \author[INFNRM]{M. Bonori},
 \author[INFNPI]{B. Bouhadef},
 \author[INFNNA,INAFNA]{M. Brescia},
 \author[INFNLNS]{G. Cacopardo},
 \author[INFNBA]{F. Cafagna},
 \author[INFNRM]{A. Capone},
 \author[INFNCT]{L. Caponetto},
 \author[INFNPI]{E. Castorina},
 \author[INFNBA]{A. Ceres},
 \author[INFNRM]{T. Chiarusi},
 \author[INFNBA]{M. Circella},
 \author[INFNLNS]{R. Cocimano},
 \author[INFNLNS]{R. Coniglione},
 \author[INFNLNF]{M. Cordelli},
 \author[INFNLNS]{M. Costa},
 \author[INFNGE]{S. Cuneo},
 \author[INFNLNS]{A. D'Amico},
 \author[INFNRM]{G. De Bonis},
 \author[INFNBA]{C. De Marzo \thanksref{Dead}},
 \author[INFNNA]{G. De Rosa},
 \author[INFNGE]{R. De Vita},
 \author[INFNLNS]{C. Distefano\corauthref{ca:fax}}
 \ead{distefano\_c@lns.infn.it},
 \author[INFNPI]{E. Falchini},
 \author[INFNBA]{C. Fiorello},
 \author[INFNPI]{V. Flaminio},
 \author[INFNGE]{K. Fratini},
 \author[INFNBO]{A. Gabrielli},
 \author[INFNPI]{S. Galeotti},
 \author[INFNBO]{E. Gandolfi},
 \author[INFNBO]{G. Giacomelli},
 \author[INFNBO]{F. Giorgi},
 \author[INFNCT]{A. Grimaldi},
 \author[INFNLNF]{R. Habel},
 \author[INFNCT,UniCT]{E. Leonora},
 \author[INFNRM]{A. Lonardo},
 \author[INFNNA]{G. Longo},
 \author[INFNCT,UniCT]{D. Lo Presti},
 \author[INFNRM]{F. Lucarelli},
 \author[INFNPI]{E. Maccioni},
 \author[INFNBO]{A. Margiotta},
 \author[INFNLNF]{A. Martini},
 \author[INFNRM]{R. Masullo},
 \author[INFNBA]{R. Megna},
 \author[INFNLNS,UniCT]{E. Migneco},
 \author[INFNBA]{M. Mongelli},
 \author[UniWi]{T. Montaruli \thanksref{UniBa}},
 \author[INFNPI]{M. Morganti},
 \author[INFNLNS]{M. Musumeci},
 \author[INFNRM]{C.A. Nicolau},
 \author[INFNLNS]{A. Orlando},
 \author[INFNGE]{M. Osipenko},
 \author[INFNNA]{G. Osteria},
 \author[INFNLNS]{R. Papaleo},
 \author[INFNLNS]{V. Pappalardo},
 \author[INFNCT,UniCT]{C. Petta},
 \author[INFNLNS]{P. Piattelli},
% \author[INFNPI]{F. Raffaelli},
 \author[INFNLNS]{G. Raia},
 \author[INFNCT]{N. Randazzo},
 \author[INFNCT]{S. Reito},
 \author[INFNGE]{G. Ricco},
 \author[INFNLNS]{G. Riccobene},
 \author[INFNGE]{M. Ripani},
 \author[INFNLNS]{A. Rovelli},
 \author[INFNBA]{M. Ruppi},
 \author[INFNCT,UniCT]{G.V. Russo},
 \author[INFNNA]{S. Russo},
 \author[INFNLNS]{P. Sapienza},
 \author[INFNLNS]{M. Sedita},
% \author[INFNRM]{J-P. Schuller \thanksref{CEA}},
 \author[INFNGE]{E. Shirokov},
 \author[INFNRM]{F. Simeone},
 \author[INFNCT,UniCT]{V. Sipala},
 \author[INFNBO]{M. Spurio},
 \author[INFNGE]{M. Taiuti},
 \author[INFNPI]{G. Terreni},
 \author[INFNLNF]{L. Trasatti},
 \author[INFNCT]{S. Urso},
 \author[INFNLNF]{V. Valente},
 \author[INFNRM]{P. Vicini},

%%%%%%%%\thanks[X]{This is the history of the paper, etc etc}

\corauth[ca:fax]{Fax: +39 095 542 398}
\thanks[Dead]{Deceased}
%\thanks[CEA]{Present address: DAPNIA/SPP Bât 141 CEN Saclay, 91191 Gif-sur-Yvette, France}
\thanks[UniBa]{On leave of absence Dipartimento Interateneo di Fisica Universit\`a di Bari, Via E. Orabona 4, 70126, Bari, Italy }

%\address[void]{ }
\address[INFNBA]{INFN Sezione Bari and Dipartimento Interateneo di Fisica Universit\`a di Bari,  Via E. Orabona 4, 70126, Bari, Italy}
\address[INFNBO]{INFN Sezione Bologna and Dipartimento di Fisica Universit\`a di Bologna, V.le Berti Pichat 6-2, 40127, Bologna, Italy}
\address[INFNCT]{INFN Sezione Catania, Via S.Sofia 64, 95123, Catania, Italy}
\address[UniCT]{Dipartimento di Fisica and Astronomia Universit\`a di Catania, Via S.Sofia 64, 95123, Catania, Italy}
\address[INFNLNS]{INFN Laboratori Nazionali del Sud, Via S.Sofia 62, 95123, Catania, Italy}
\address[INFNLNF]{INFN Laboratori Nazionali di Frascati, Via Enrico Fermi 40, 00044, Frascati (RM), Italy}
\address[INFNGE]{INFN Sezione Genova and Dipartimento di Fisica Universit\`a di Genova, Via Dodecaneso 33, 16146, Genova, Italy}
\address[INFNNA]{INFN Sezione Napoli and Dipartimento di Scienze Fisiche Universit\`a di Napoli, Via Cintia, 80126, Napoli, Italy}
\address[INAFNA]{INAF Osservatorio Astronomico di Capodimonte, Salita Moiariello 16, 80131, Napoli, Italy}
\address[INFNPI]{INFN Sezione Pisa and Dipartimento di Fisica Universit\`a di Pisa, Polo Fibonacci, Largo Bruno Pontecorvo 3, 56127, Pisa, Italy}
\address[INFNRM]{INFN Sezione Roma 1 and Dipartimento di Fisica Universit\`a di Roma "La Sapienza", P.le A. Moro 2, 00185, Roma, Italy}
\address[UniWi]{University of Wisconsin, Department of Physics, 53711, Madison, WI, USA}

\begin{abstract}
% Text of abstract

In this paper are presented the results of Monte Carlo simulations
on the capability of the proposed NEMO-km$^3$ telescope to detect
TeV muon neutrinos from Galactic microquasars. For each known
microquasar we compute the number of detectable events, together
with the atmospheric neutrino and muon background events. We also
discuss the detector sensitivity to neutrino fluxes expected from
known microquasars, optimizing the event selection also to reject
the background; the number of events surviving the event selection
are given. The best candidates are the steady microquasars SS433
and GX339-4 for which we estimate a sensitivity of about
$5\cdot10^{-11}$ erg/cm$^2$ s; the predicted 
fluxes are expected to be well above
this sensitivity. For bursting microquasars the most interesting
candidates are Cygnus X-3, GRO J1655-40 and XTE J1118+480: their
analyses are more complicated because of the stochastic nature of
the bursts.

\end{abstract}

\begin{keyword}
Microquasars \sep Neutrino telescopes \sep NEMO
\PACS 95.55.Vj \sep %Neutrino, muon, pion, and other elementary particle detectors; cosmic ray detectors
95.85.Ry \sep %Neutrino, muon, pion, and other elementary particles; cosmic rays
96.40.Tv  %Neutrinos and muons
\end{keyword}
\end{frontmatter}

% main text
\section{Introduction}
\label{sec:introduction}

The realization of a km$^3$ scale detector for astrophysical high energy neutrinos is
one of the most important scientific goals of astroparticle physics.
Due to their small interaction cross section,
high energy neutrinos are expected to allow us to
observe inside dense environments in our Universe and to
extend our knowledge beyond the distances that can be
explored with $\gamma$-rays and cosmic ray observatories
\cite{Learned00,Halzen02}.

A number of astrophysical sources of high energy neutrinos, both galactic
(SuperNova remnants, microquasars, ...) and extragalactic (Active Galactic Nuclei,
Gamma Ray Bursts, ...), have been proposed.
Since neutrino
production requires a hadronic component inside the jet, neutrino
detection will be a fundamental probe to investigate jet
composition and particle acceleration processes taking place in these sources.
%Neutrinos can also probe the
%source smallest scales that are dense and opaque to
%electromagnetic radiation.

The main aim of this paper is to study the sensitivity of the proposed NEMO-km$^3$
telescope \cite{nemo} to high energy neutrinos emitted by known microquasars.
In particular we simulated the interaction of microquasar muon-neutrinos
inside and in the proximity of the NEMO-km$^3$ neutrino telescope in order
to estimate the expected muon event rate.
In our calculations we accounted for the atmospheric neutrino and muon
fluxes reaching the detector in order to optimize the event selection and
reconstruction procedures.

\section{Neutrino fluxes from microquasars}

Microquasars are Galactic X-ray binary systems which exhibit
relativistic jets, observed in the radio band. The presence of an
accretion disc and relativistic jets make microquasars similar to
{\it small} quasars (see ref. \cite{Chaty05}).
Several authors propose microquasar jets as sites of acceleration of charged particles
up to energies of about $10^{16}$ eV, and of high energy neutrino production.

Levinson \& Waxman \cite{Levinson01} proposed
a theoretical model for neutrino emission from microquasar jets,
considering the production of neutrinos from the interactions of protons, accelerated
in the jet, with synchrotron photons, emitted by electrons, or with
external X-ray photons of the accretion disk.
According to this model, microquasars may be intense sources of high energy muon
neutrinos, in the energy range between 1 and 100 TeV.
Applying the Levinson \& Waxman model, Distefano et al. \cite{Distefano02}
estimated the expected neutrino fluxes from an ensemble of known microquasars,
showing that several of these sources may be detected with the future km$^3$ neutrino
telescopes.

Bednarek \cite{Bednarek05} considers the case of microquasars in
which a Wolf-Rayet star supplies matter onto the compact object.
Together with the scenario proposed by Levinson and Waxman,
Bednarek considers also two regions where high energy neutrinos
may be produced. In the jet region II (see ref.\cite{Bednarek05})
neutrons resulting from photo-disintegration of accelerated nuclei
produce neutrinos after interactions with the inner disk matter.
In the jet region III the process takes place in the proximity of
the Wolf-Rayet star surface. In particular Bednarek discusses the
case of Cygnus X-3 and calculates the neutrino spectra (see Fig. 2
in \cite{Bednarek05}) produced by a power-law spectrum of nuclei
$\gamma_A^{-\kappa}$, exploring different possible values for the
spectral index $\kappa$ and for the Lorentz factor cut-off
$\gamma_A^{max}$. Another component of the neutrino flux may arise
from neutrons injected by mono-energetic nuclei having a Lorentz
factor $\gamma_A^{min}\approx10^5$.

Aharonian et al. \cite{Aharonian05} discuss different possible scenarios for the production
of the $\gamma$-ray flux recently observed from LS 5039 by the HESS Collaboration \cite{hess}.
They consider both leptonic and hadronic production mechanisms and argue
in favor of a TeV photon flux originating from pp interactions.
They suggest therefore that the $\gamma$-ray flux should be accompanied by a TeV neutrino flux
of about $10^{-12}$ cm$^{-2}$ s $^{-1}$, or even a factor of 100 larger.

In this paper we discuss the results obtained simulating neutrino fluxes calculated
with the three models reported above.

Christiansen et al. \cite{Christiansen05} recently published
predictions on the emission of high energy neutrinos from ``windy
microquasars". Their calculations, based on the model proposed by
Romero et al. \cite{Romero03}, describe the emission of high
energy neutrinos and $\gamma$-rays from pions created in the
inelastic collisions between relativistic protons ejected by the
compact object and ions in the stellar wind. As an example they
consider microquasar LS I +61 303, estimating an average event
rate of 3-5 TeV muon neutrinos per kilometer-square per year.
%This
%source is outside the field of view of a Mediterranean neutrino
%telescope, but their model can be applied to other possible
%sources such as LS 5039, Cygnus X-3, and Cygnus X-1, which are
%observable from the Northern Hemisphere.
This source is outside the field of view of a Mediterranean
neutrino telescope and it is not considered in this work.

Romero \& Orellana \cite{Romero05} discuss the case of
microquasars that show strongly misaligned jets with respect to the perpendicular to the orbital plane.
If the donor star is an early-type star, the jet could collide with the stellar wind,
producing $\gamma$-rays and neutrinos. The expected neutrino fluxes are in general too
low to be detected by the planned km$^3$ telescopes and are not considered in this paper.

\section{The NEMO project}

The NEMO Collaboration  \cite{VLVnT2} is performing
R\&D towards the design and construction of the Mediterranean
km$^3$ neutrino detector.
The activity was mainly focused on the search and characterization of an
optimal site for the detector installation and on the
development of a feasibility study for the detector.

After eight years of activities in seeking and monitoring marine sites in the central Mediterranean Sea,
the Collaboration selected a large marine area (centered at the coordinates Lat. $36^\circ$ 25' N, Long. $16^\circ$ 00' E)
located about 80 km from the Southern cape of Sicily, Capo Passero, in the Ionian Sea plateau
as an optimal site for the deployment of the km$^3$ telescope.
The measured average values of the absorption length for blue light (440 nm) is  $\sim70$ m, close
to the one of optically pure sea salt water; measurements also show that seasonal variations of the light absorption length
are negligible \cite{Riccobene06}.
The optical background, measured at $\sim3000$ m depth
and for a 10" PMT at 0.5 s.p.e. threshold, has an average rate of 30 kHz,
compatible with the noise expected from $^{40}$K decay.
An exhaustive report on the Capo Passero site properties is available in \cite{Riccobene06,appec}.

NEMO proposes a preliminary project for the km$^3$ detector based on present technological and budget constraints.
The detector architecture consists of a
square array of structures called  towers, hosting the optical modules and the instrumentation.
The tower is a three dimensional flexible structure composed of a
sequence of ``storeys" (which host the instrumentation) interlinked by a system of cables and anchored on the seabed. The
structure is kept vertical by an appropriate buoyancy on the top.
In its working position
each storey will be rotated by 90$^\circ$, with respect to the upper and lower adjacent ones, around the vertical axis of the tower.
The final features of the tower (number and length of storeys, number of optical modules per storey, distance between the storeys,
distance between the towers) is under study with the goal of optimizing the detector performance.

The km$^3$ telescope, simulated in this paper, is a square array of
$9\times9$ towers with a distance between towers of 140 m.
In this configuration each tower hosts 72 PMTs (with a diameter of 10"), namely 5832 PMTs for the
whole detector with a total geometrical volume of $\sim0.9$ km$^3$.
We considered an 18 storey tower; each storey
is made of a 20 m long beam structure hosting two optical modules (one downlooking and one looking horizontally) at each end
(four OMs per storey). The vertical distance between storeys is 40 m. A spacing of 150 m is added at the base of the tower,
between the anchor and the lowermost storey.

As an intermediate step to ensure an adequate process of
validation, NEMO built a demonstrator which includes most of the
critical elements of the proposed km$^3$ detector: NEMO Phase-1
(\cite{VLVnT2} and ref. therein). It was realizated at the
Underwater Test Site of the Laboratori Nazionali del Sud in
Catania, where a 28 km electro optical cable, reaching the depth
of 2000 m, allows the connection of deep sea intrumentation to a
shore station. The NEMO Phase-1 system is composed of a junction
box and a mini-tower (with 4 storeys). This allows to test the
proposed mechanical solutions as well as the data transmission,
the power distribution, the timing calibration and the acoustic
positioning systems. NEMO Phase-1 is operating since December 2006
\cite{rosa}.

Although the Phase-1 project will provide a fundamental test of the technologies proposed for the realization and installation
of the detector, these must be finally validated at the depths needed for the km$^3$ detector. For these motivations the
realization of an infrastructure on the site of Capo Passero has been undertaken \cite{VLVnT2}.

A further R\&D program will also be developed by the Collaboration within the KM3NeT Design Study \cite{km3net}.

\section{The simulation codes}

In this section we describe the Monte Carlo (MC) simulation tools
and procedures to study the detector response to high energy neutrinos.
As a first step we generated samples of muon events having vertexes
within an appropriate volume of water.
Three sets of muon samples were generated:
atmospheric muons, muons induced by atmospheric neutrinos and
muons induced by neutrinos coming from point-like astrophysical sources.
These events were then propagated inside the detector,
using the simulation tools developed by the
ANTARES Collaboration \cite{ANTARES}
described in ref. \cite{Becherini05}.
The codes simulate the emission and the propagation
of \v{C}erenkov light radiated by muons and their secondary
products, then record photo-electrons signals on PMTs.
Subsequently, background signals due to $^{40}$K decay
are generated in an appropriate time window and added to the event.
Another code \cite{Becherini05} is then
used to reconstruct muon tracks.
These codes, developed for detector configurations with smaller size and different geometries,
were modified for a larger geometry, as described in \cite{Zaborov02,Sapienza03}.
A brief description of the simulation steps is given in
the following.

\subsection{Generation of neutrino-induced muon events at the detector}

Neutrino-induced muon events are generated using the
event generation code developed by the ANTARES Collaboration \cite{Becherini05}.
The code generates
neutrinos that interact inside and in the proximity of the detector,
producing detectable muons.
Interacting neutrinos are generated with a
power law spectrum $\varepsilon_\nu^{-X}$, where $X$ is
a generic spectral index.
These events are weighted, as described below,
in order to reproduce the expected spectrum.
The direction of incident neutrinos coming from a point-like source
is generated taking into account the change of the relative position of the
astrophysical sources with respect to the detector rest frame,
due to the Earth rotation. In the case of a diffuse
neutrino flux, such as atmospheric neutrinos,
the neutrino direction is generated with an isotropic distribution.

Once the interacting neutrino is generated (i.e. neutrino energy $\varepsilon_\nu$,
direction ($\vartheta_\nu,\varphi_\nu$), interaction vertex ($x,y,z$)
and event weight $W_{event}$ are assigned), the code simulates
the neutrino Charged Current weak interaction, producing the muon kinematics.
Since the detection efficiency of the
detector is poor for muon energies lower than 100 GeV \cite{Sapienza03},
the lowest neutrino energy considered in the simulations is 100 GeV.
At these energies, the main contribution to the neutrino interaction, is from
Deep Inelastic Scattering (DIS), here simulated using the code LEPTO \cite{Ingelman96}.
Once the neutrino-induced muons are generated, they are
propagated up to the surface of the detector sensitive volume.
This volume is defined as a cylinder with size large enough that the
\v{C}erenkov light emitted outside its volume has a negligible
probability to reach any PMT. In particular, the cylinder size extends beyond the instrumented
volume by 3 times the blue light absorption length ($\sim 70$ m).
The muon propagation is performed using the code MUSIC
\cite{Antonioli97}.

In the generation procedure a weight $W_{event}$ is assigned
to each event in order to normalize the number of generated events to the
expected neutrino flux.
The event weight is calculated multiplying the expected neutrino
spectrum $({dn_\nu/d\varepsilon_\nu dSdt})^{expected}$
(evaluated at the generated neutrino energy and direction) times the
generation weight $W_{gen}$:
\begin{equation}
W_{event}= W_{gen}\cdot \left(\frac{dn_\nu}{d\varepsilon_\nu
dSdt}\right)^{expected}.
\label{eq:weight_event1}
\end{equation}
$W_{gen}$ is defined as the inverse of the simulated
neutrino spectrum. In the case of a diffuse neutrino flux, it is given by:
\begin{equation}
%\begin{array}{cl}
W_{gen}^{-1} =\displaystyle{\frac{\varepsilon_\nu^{-X}}{I_E I_\vartheta}}\cdot
\frac{N_{total}}{t_{gen}}\cdot\frac{1}{V_{gen}} \cdot
\frac{1}{\sigma_{CC}(\varepsilon_\nu)\rho N_A}\cdot
\frac{1}{P_{Earth}(\varepsilon_\nu,\vartheta_\nu)},
%\end{array}
\label{eq:spectrum_gen}
\end{equation}
where $X$ is the generation spectral index, $I_E$ is the
integral of the generation spectrum shape $\varepsilon_\nu^{-X}$
over the whole simulated neutrino energy range,
$I_\vartheta$ is the integral of the solid angle in which events are generated.
$N_{total}$ is the number of simulated events and $t_{gen}$ is the
event generation time. The neutrino interaction vertices
are randomly generated within the volume $V_{gen}$ that
completely contains the detector sensitive volume and with size large
enough that neutrinos interacting outside cannot produce
detectable muons. In particular, its size extends beyond the
detector sensitive volume by the maximum range covered by the
neutrino-induced muons.
$(\sigma_{CC}(\varepsilon_\nu)\rho N_A)^{-1}$
is the neutrino CC interaction length in a medium of density
$\rho$, $\sigma_{CC}(\varepsilon_\nu)$ is the CC neutrino interaction cross
section and $N_A$ is the Avogadro number.
Eq.(\ref{eq:spectrum_gen}) takes also into account the neutrino
absorption in the Earth, through the transmission
probability $P_{Earth}(\varepsilon_\nu,\vartheta_\nu)$,
which depends on neutrino energy and direction \cite{Gandhi95}.
In the case of a point-like source neutrinos,
the factor $I_\vartheta$ does not appear in relation \ref{eq:spectrum_gen}.

\subsection{Generation of atmospheric muons}
\label{sec:atmuon}

Atmospheric muons are generated according to the
Okada formula \cite{Okada94}, which parameterizes
the integral energy spectrum of muons, reaching a given
depth $D$ in seawater, as a function of the muon energy
and direction. Since this parameterization is a complicated
function of many parameters, a
weighted generation procedure, similar to the one
used for neutrino events, is used.
Muon energies are generated with a given spectral index $X$
and muon directions are isotropically generated, in $2\pi$ solid angle
(only downward directed). For each generated
event a weight $W_{event}$ is calculated multiplying the Okada
atmospheric muon spectrum times the generation
weight $W_{gen}$
\begin{equation}
W_{gen}=\frac{\varepsilon_\mu^X I_E I_\vartheta t_{gen}
A_{geom}}{N_{total}},
\end{equation}
where $A_{geom}$ is the detection geometrical area, i.e. the
detector sensitive surface, projected on a plane perpendicular
to the incident muon direction.
The other parameters have the same meaning
as in Eq.(\ref{eq:spectrum_gen}).
In order to take into account the detector height ($\simeq 1$ km),
the Okada spectrum is evaluated at the depth where the
muon track intersects the detector sensitive volume.

\subsection{Muon propagation inside the detector}

The subsequent steps of the simulation are: the propagation of
muons inside the detector, the generation of the \v{C}erenkov
photons and the simulation of the PMT response. The muon
propagation is performed with MUSIC  \cite{Antonioli97} in order
to take into account high energy radiative processes like
bremsstrahlung. The \v{C}erenkov light produced by secondary
electrons is also taken into account. In the code we consider
the light absorption length as a function of wavelength,
according to the water optical properties measured in the Capo
Passero site. Once the PMT hits are generated, spurious PMT hits,
due to the underwater optical noise ($^{40}$K decays), are
introduced. It is assumed that optical background produces
uncorrelated single photo-electron (s.p.e.) signals in the PMTs.
The PMT response function is taken into account. In the
simulations made for this work the optical background signals are
generated assuming an average rate of 30 kHz for 10" PMTs,
corresponding to the average value measured in Capo Passero.

\subsection{Event reconstruction}

It is possible to reconstruct the direction of the muon track,
using information from arrival times of photons in the
PMTs, hit amplitudes (proportional to the number of photo-electrons in the PMTs) and PMT positions.
The used reconstruction strategy aims at the
rejection of hits due
to background and selects a sub-set of hits compatible with the muon track. The track reconstruction algorithm used in this
work is a robust track fitting procedure based on a maximum likelihood method \cite{Heijboer03}.

In order to reduce the number of hits due to background the first step consists in the rejection of hits with amplitude
smaller than 0.5 p.e. then a causality filter with respect to the highest amplitude hit is applied\footnote{This is an
important step since the number of noise hits may be comparable or
larger than the number of muon hits in case of high noise rate or low
muon energy.}. The causality filter
used in this work is based on the following
condition:
\begin{equation}
\label{eq:causality}
(|dt|-dr/v<20\hbox{ ns}) \hbox{ AND } (||dt|-dr/c|<500\hbox{ ns}),
\end{equation}
where $|dt|$ is the absolute value of the time delay
between two hits, $dr$ is the distance between the PMTs where the
hits are detected, $v$ is the group velocity of light in water.
The first condition is based on the fact that the front
propagation of direct \v{C}erenkov photons moves in water with
velocity $v \approx c/n$ ($n=1.35$). The second condition
takes into account that, in the case of large distances,
light absorption does not allow \v{C}erenkov photons to propagate
far from the initial muon track. The value of 500 ns
takes into account the late \v{C}erenkov photons, due to
light scattering, $\delta$-rays and showers.

The reconstruction algorithm starts with a linear prefit applied on a sub-set of
hits that passed the causality criterion and in which there are at least three
hits that satisfy the following trigger
condition: the hit has an amplitude higher than 3.5 p.e. or the hit belongs
to a ``local coincidence". A local coincidence is defined as two or more simultaneous
hits ($\Delta t\leq 20$ ns) in the two PMTs at the bar edge.
Starting from the result of the pre-fit, a sequence of fit procedures using all the hits
that passed the causality criterion is applied. A further hit rejection based on
the track parameters is also applied.

\subsection{Criteria for atmospheric background rejection}

At the reconstruction level the fraction of background events is very high. In order to reduce it,
an event selection is applied.
The used reconstruction algorithm is a robust track fitting procedure based
on a maximization likelihood method \cite{Heijboer03}.
In this work, we used, as a {\it goodness of fit criterion}, the variable:
\begin{equation}
\Lambda\equiv -\frac{\hbox{log}(\mathcal{L})}{N_{DOF}}+0.1(N_{comp}-1),
\end{equation}
where $\hbox{log}(\mathcal{L})/N_{DOF}$ is the log-likelihood
per degree of freedom ($N_{DOF}$) and $N_{comp}$ is the number of
solutions found by the reconstruction program
(see ref. \cite{Heijboer03} for details). This quality cut is
applied together with the selection criteria listed here:

\begin{itemize}

\item the muon must be reconstructed with a Zenith angle $\vartheta_\mu^{rec}<\vartheta_\mu^{max}$,
in order to reject down-going events;

\item the variable $\Lambda$ must be greater than a given value
$\Lambda_{cut}$;

\item only events reconstructed in a circular sky region
centered on the source position and having a radius of $r_{bin}$
are considered.

\end{itemize}

\section{NEMO-km$^3$ sensitivity to microquasars neutrino fluxes}
\label{sec:sensitivity}

The detector sensitivity was calculated according to the Feldman and Cousins
approach \cite{Feldman98}. The 90\% c.l. sensitivity to a neutrino flux coming from a point-like source
is given by
\begin{equation}
f_{\nu,90}=\frac{\overline{\mu}_{90}(b)}{N_\mu^m}f_{\nu}^{th},
\end{equation}
where $\overline{\mu}_{90}(b)$ is the 90\% c.l. average upper
limit for an expected background (atmospheric neutrinos + muons)
with known mean value $b$ and no true signal \cite{Feldman98},
$f_{\nu}^{th}$ is the theoretical neutrino flux from a given microquasar that induces
a mean signal $N_\mu^m$. During the calculation of the
sensitivity, an event selection is applied as described in the
previous section. The optimal values of $\vartheta_\mu^{max}$,
$\Lambda_{cut}$ and $r_{bin}$ are chosen to optimize the detector
sensitivity.

\subsection{Atmospheric neutrino background simulation}

In order to evaluate the contribution of the atmospheric neutrino
background with respect to the signal produced by microquasars, we
simulated a total number $N_{total}=5.1\cdot10^{10}$ interacting
atmospheric neutrinos, with a generation spectral index $X=2$.
Neutrinos were generated in the energy range 100 GeV - $10^8$ GeV
with a $4\pi$ isotropic angular distribution. The corresponding
total number of reconstructed neutrino-induced muons is 
$\sim6.6\cdot 10^5$. The atmospheric neutrino events are weighted to
the sum of the Bartol flux \cite{Agrawal96} and of the {\tt rqpm}
model \cite{Bugaev98} to take into accunt the contribution of
prompt neutrinos. Considering an observation time of 1 year, the
total weighted number of reconstructed atmospheric neutrinos
(up-going + down-going) is $\sim4.7\cdot 10^4$ (quality cuts are
not yet applied).

\subsection{Atmospheric muon background simulation}

Atmospheric muon events are also generated applying a weighted
generation technique. We generated $N_{total}= 2.5\cdot 10^7$
muons, in the energy range 1 TeV - 1 PeV, with a spectral index
$X=3$ and $N_{total}= 4\cdot 10^7$ muons in the range 100 GeV - 1
TeV, with a spectral index $X=1$. Muons are generated with an
isotropic distribution as described in sec. \ref{sec:atmuon}. The
corresponding number of reconstructed events is 
$\sim9\cdot10^5$ in the range 100 GeV - 1 TeV and $3.1\cdot10^6$ in
the range 1 TeV - 1 PeV (in total $\sim 4\cdot 10^6$).

According to the Okada parameterization, the expected number of
reconstructed muons is $\sim1.3\cdot 10^9$ per year. The
statistics of the generated muon sample corresponds to only a few
days of data taking. Since reconstructed events have a flat
distribution in Right Ascension (RA), it is possible to project
the full sample of simulated events in a few degrees bin
$\Delta$RA. This allows to obtain statistics of atmospheric muon
background corresponding to a time of $\sim1$ year of data taking
for a region of a few degrees around each single source.% (the
%exact detector data taking time interval is a function of the
%bin-width $\Delta$RA and of the declination).

\subsection{Simulation of neutrino events from microquasars}

For each source, a number $N_{total}=10^9$ of interacting
neutrinos was simulated in the energy range 1 - 100 TeV
in order to reproduce the neutrino fluxes calculated by Distefano et al.
\cite{Distefano02}.
The generation spectral index is chosen to be $X=1$, in order to guarantee
a good event statistics at the highest energies.
For microquasar LS 5039, $N_{total}=10^8$ events were generated
in the energy range 0.1 - 1 TeV to simulate neutrinos events
according to the model proposed by Aharonian et al. \cite{Aharonian05}.
For microquasar Cygnus X-3, $N_{total}=10^8$ events were generated
in the energy range 0.1 - 1 TeV and $N_{total}=10^9$ events with energies 100 - 1000 TeV
to simulate Bednarek's theoretical fluxes \cite{Bednarek05}.

%\subsection{``Neutrino image" of a point-like source}

Using the information on
reconstructed muon directions we computed the equatorial
coordinates (declination and right ascension) of the detected muons,
and we mapped the reconstructed events in equatorial
coordinates.
This allows to obtain an image of the source and to evaluate the spread of the reconstructed muon
directions (due to neutrino interactions and to the muon
track reconstruction error) around the source coordinates.

\begin{figure}[h]
\begin{center}
\includegraphics[width=7.5cm]{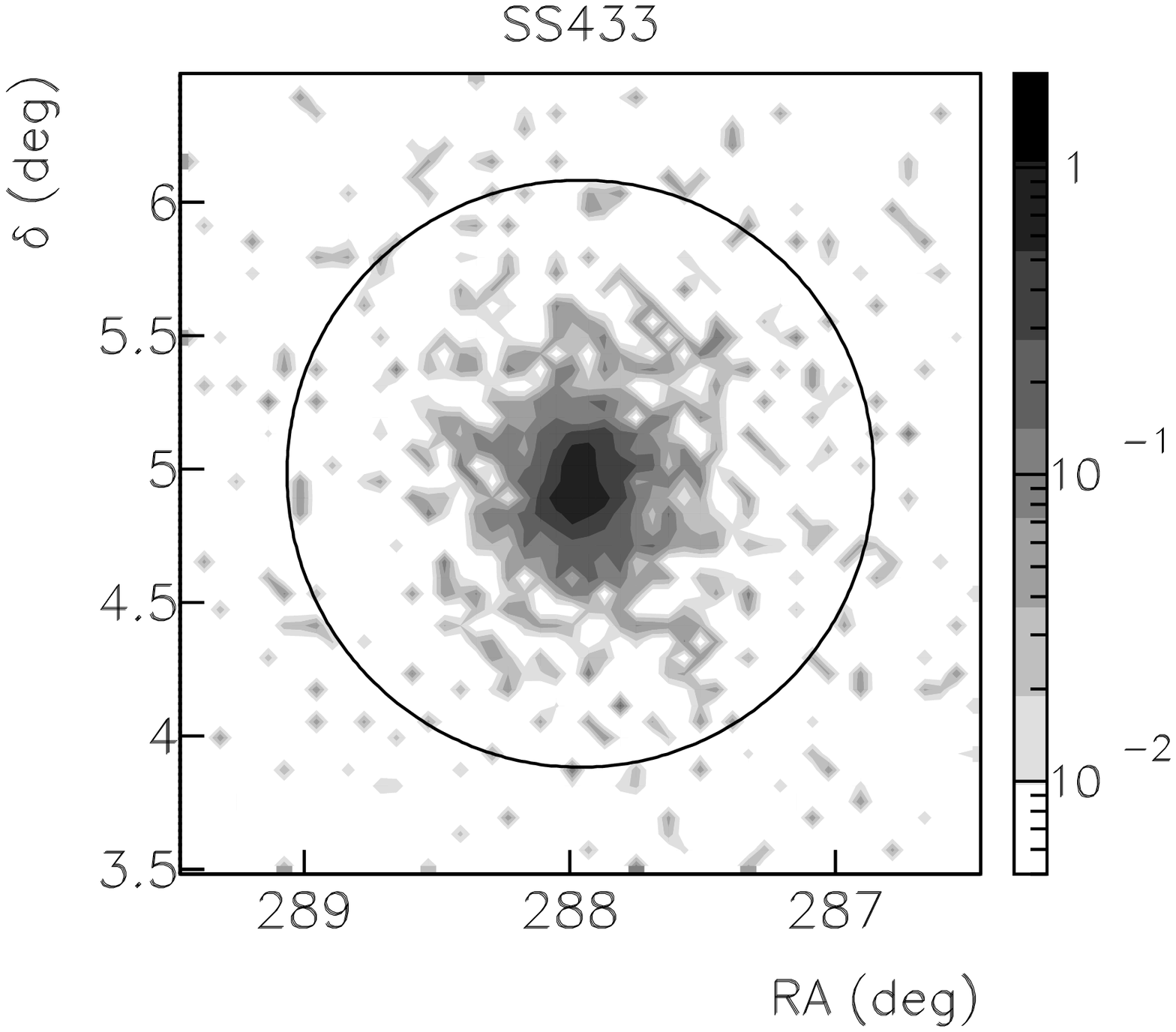}\includegraphics[width=7.5cm]{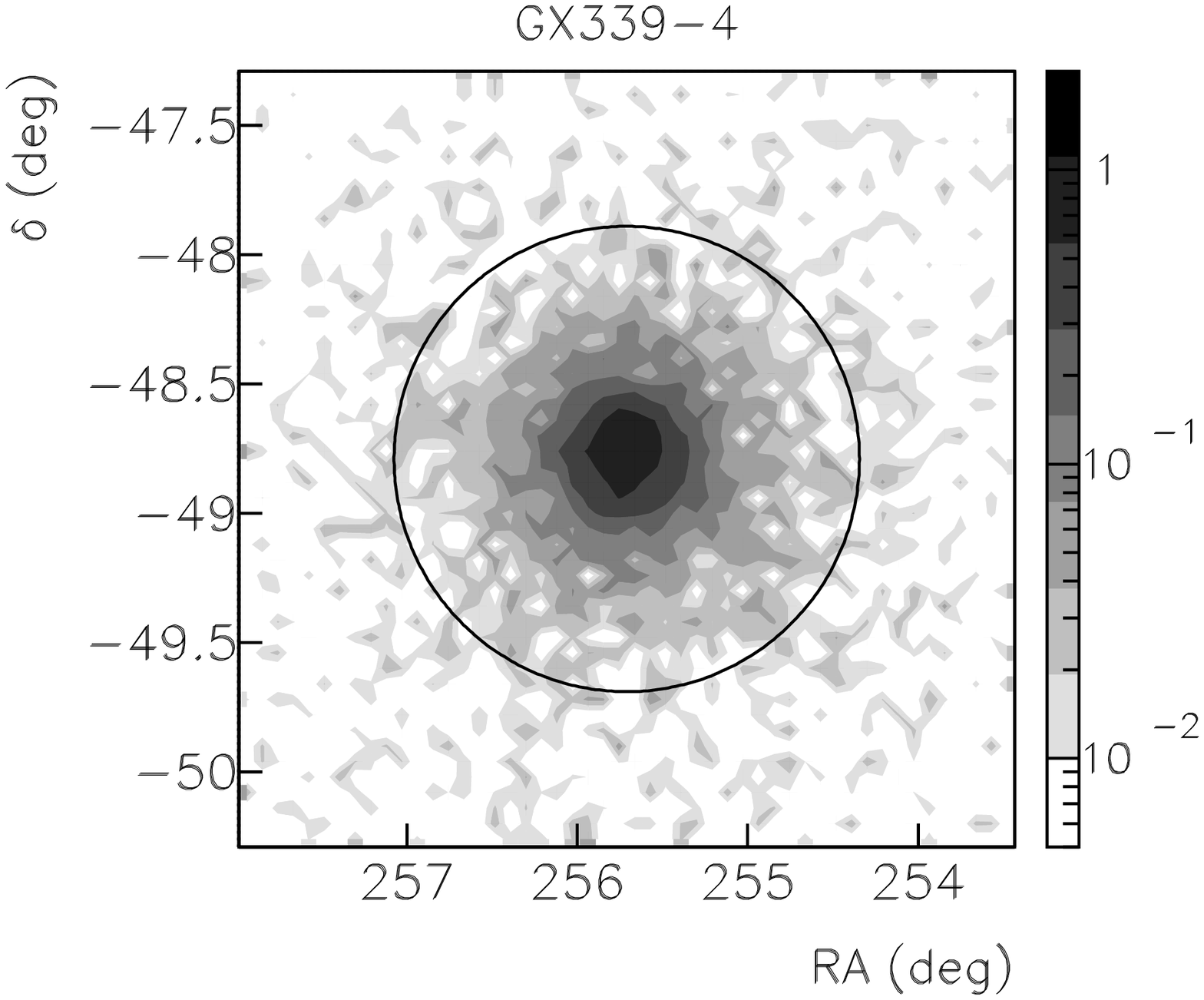}
\end{center}
\vspace*{-0.5cm} \caption{Weighted MC event distributions for the
microquasars SS433 and GX339-4, referring to 1 year of data
taking. Both maps are obtained from 2D histograms with
$0.06^\circ\times0.06^\circ$ bin size. The circle has a radius
equal to the median of the angular spread distribution of the
reconstructed tracks. It represents the region where 50\% of the
events are reconstructed. } \label{fig:mappe}
\end{figure}

The 2D angular distribution of the reconstructed muons for SS433
and GX339-4 (the two persistent microquasars expected to produce
the largest number of events) is plotted in Fig. \ref{fig:mappe}.
The angular resolution is calculated as the median of the angular
spread distribution of the reconstructed tracks with respect to
the source direction. It is 1.1$^\circ$ for SS433 and 0.9$^\circ$
for GX339-4 (see Fig. \ref{fig:mappe}). The angular resolution
value depends on the source theoretical spectral index
($\Gamma=2$).% and on the energy range of the simulated microquasar
%neutrino flux.

\subsection{Detector sensitivity to neutrino microquasars}

The detector sensitivity was calculated for a livetime of 1 year,
simulating a neutrino flux with spectral index $\Gamma=2$ in the
energy range 1 - 100 TeV. The study was carried out for each
microquasar, since the sensitivity is a function of the source
astronomical declination.

\begin{figure}[h]
\begin{center}
\includegraphics[width=9.4cm]{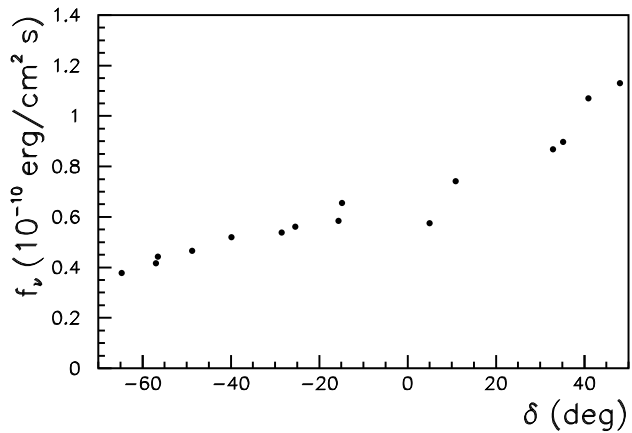}
\end{center}
\caption{NEMO-km$^3$ sensitivity to neutrinos from microquasars versus source declination,
for a livetime of 1 year. The worsening of the sensitivity with increasing declination is
due to the decrease of the source visibility.}
\label{fig:sensi-ene-flusso-delta-1years}
\end{figure}

\begin{figure}[h]
\begin{center}
\includegraphics[width=9.4cm]{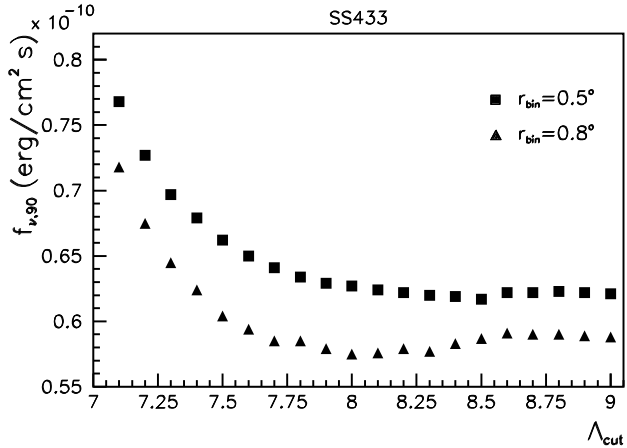}
\end{center}
\caption{NEMO expected sensitivity for microquasar SS433 for the applied
$\vartheta_\mu^{max}$ and $r_{bin}$ values as a function of the
quality cut $\Lambda_{cut}$.} \label{fig:sensi-cut}
\end{figure}

Tab. \ref{tab:sensitivity} gives the detector sensitivity
$f_{\nu,90}$ for each microquasar and the corresponding values of
$\vartheta_\mu^{max}$, $\Lambda_{cut}$ and $r_{bin}$ which
optimize $f_{\nu,90}$. Fig.
\ref{fig:sensi-ene-flusso-delta-1years} shows the detector
sensitivity for the studied microquasars as a function of the
declination; the sensitivity flux limit increases with increasing
declination, due to the decrease of the time per day spent by the
source below the Astronomical Horizon (with respect to the
latitude of the Capo Passero site). In Tab. \ref{tab:sensitivity},
we present only results concerning microquasars which may be
observed by a telescope located in the Capo Passero site (i.e.
with declination $\delta\ltsim+54^\circ$). For the most
interesting steady microquasars the sensitivity is
$\sim5\cdot10^{-11}$ erg/cm$^2$ s, see Fig. \ref{fig:sensi-cut}.

In Tab. \ref{tab:sensitivity_aharonian} are summarized the
detector sensitivities for microquasar LS 5039, according to the
model of Aharonian et al. \cite{Aharonian05}. The simulations
refer to a power-law neutrino spectrum,
$dn_\nu/d\varepsilon_\nu\propto\varepsilon_\nu^{-\Gamma}$, with
energy cutoff $\varepsilon_\nu^{max}=10$ TeV and 100 TeV, and
$\Gamma=1.5$ and 2, respectively. For the four combinations of
parameters $\Gamma$ and $\varepsilon_\nu^{max}$, an average
neutrino energy flux $f_\nu^{th}\sim10^{-10}$ erg/cm$^2$ s
($\varepsilon_\nu>0.1$ TeV) was simulated.

\begin{table}[h]
\renewcommand{\arraystretch}{1.2}
\begin{center}
\caption{{\bf Detector sensitivity to neutrinos from microquasars:}
the sensitivity $f_{\nu,90}$ is calculated for an $\varepsilon_\nu^{-2}$
neutrino spectrum in the energy range 1 - 100 TeV, for a detector livetime 
of 1 year. The corresponding values of $\vartheta_\mu^{max}$, $\Lambda_{cut}$ and $r_{bin}$
and the source declination $\delta$ are also given.}
\label{tab:sensitivity}
~\\*[0.2cm]
\begin{tabular}{lccccc}
\hline
     {\bf Source name}  & {\bf ${\bf \vartheta_\mu^{max}}$  (deg)}   & {\bf ${\bf \Lambda_{cut}}$} & {\bf ${\bf r_{bin}}$ (deg)} & {\bf ${\bf f_{\nu,90}}$ (erg/cm$^2$ s)}  & {\bf $\delta$ (deg)} \\
\hline
\hline
{\it Steady Sources} \\
\hline
  LS 5039                 &  101    &   -7.3          & 0.9     & $6.5\cdot10^{-11}$    & -14.85    \\
  Scorpius X-1            &  104    &   -7.7          & 0.7 	& $5.8\cdot10^{-11}$    & -15.64    \\
  SS433                   &  115    &   -8.0          & 0.8 	& $5.7\cdot10^{-11}$    & ~~4.98    \\
  GX 339-4                &  ~96    &   -7.4          & 0.5 	& $4.7\cdot10^{-11}$    & -48.79    \\
  Cygnus X-1              &  103    &   -7.5          & 0.7 	& $9.0\cdot10^{-11}$    & ~35.20    \\
\hline
{\it Bursting Sources} \\
\hline
  XTE J1748-288           &  102    &   -7.6          & 0.9     & $5.4\cdot10^{-11}$    & -28.47    \\
  Cygnus X-3              &  101    &   -7.3          & 0.8     & $1.1\cdot10^{-10}$    & ~40.95    \\
  GRO J1655-40            &  101    &   -7.4          & 0.7     & $5.2\cdot10^{-11}$    & -39.85    \\
  GRS 1915+105            &  100    &   -7.4          & 0.8     & $7.4\cdot10^{-11}$    & ~10.86    \\
  Circinus X-1            &  ~90    &   -7.3          & 0.9     & $4.2\cdot10^{-11}$    & -56.99    \\
  XTE J1550-564           &  ~90    &   -7.1          & 0.9     & $4.4\cdot10^{-11}$    & -56.48    \\
  V4641 Sgr               &  102    &   -7.4          & 0.9 	& $5.6\cdot10^{-11}$    & -25.43    \\
  GS 1354-64              &  ~90    &   -7.5          & 1.0     & $3.8\cdot10^{-11}$    & -64.73    \\
  GRO J0422+32            &  103    &   -7.5          & 0.8 	& $8.7\cdot10^{-11}$    & ~32.91    \\
  XTE J1118+480           &  102    &   -7.5          & 0.7     & $1.1\cdot10^{-10}$    & ~48.05    \\
\hline
\end{tabular}
\\*[1.cm]
\end{center}
\end{table}

\begin{table}[h]
\renewcommand{\arraystretch}{1.2}
\begin{center}
\caption{{\bf  Detector sensitivity flux to neutrinos from the steady source LS 5039:}
the sensitivity $f_{\nu,90}$ is calculated for a $\varepsilon_\nu^{-\Gamma}$
neutrino spectrum in the energy range from 0.1 TeV up to $\varepsilon_\nu^{max}$ and for a detector live
time of 1 year. The corresponding values of $\vartheta_\mu^{max}$, $\Lambda_{cut}$ and $r_{bin}$
are also given.}
\label{tab:sensitivity_aharonian}
~\\*[0.2cm]
\begin{tabular}{lccccc}
\hline
\hline
     {\bf ${\bf \Gamma}$} & {\bf ${\bf \varepsilon_\nu^{max}}$ (TeV)} & {\bf ${\bf \vartheta_\mu^{max}}$  (deg)}   & {\bf ${\bf \Lambda_{cut}}$} & {\bf ${\bf r_{bin}}$     (deg)} & {\bf ${\bf f_{\nu,90}}$ (erg/cm$^2$ s)}  \\
\hline
  1.5            & ~10      & 101      &  -7.3      & 1.0  & $1.5\cdot10^{-10}$   	\\
  1.5            & 100      & 101      &  -7.3      & 0.9  & $5.2\cdot10^{-11}$    	\\
  2.0            & ~10      & ~97      &  -7.4      & 1.0  & $2.7\cdot10^{-10}$   	\\
  2.0            & 100      & 101      &  -7.3      & 0.9  & $9.6\cdot10^{-11}$    	\\
\hline
\end{tabular}
\\*[1.cm]
\end{center}
\end{table}

\section{Expected number of microquasar events}

In Tab. \ref{tab:N_mu_simulated} are given the number of selected
neutrino events from each microquasar, applying the cuts reported
in Tab. \ref{tab:sensitivity} and according to the neutrino fluxes
given by Distefano et al. \cite{Distefano02}. The results refer to an
integration time $\Delta t$ equal to the duration of the
considered burst for the transient sources and to one year for the
steady sources. In the same table is given the background
(atmospheric neutrinos + muons) in 1 year of data taking. In this
analysis, it is assumed that transient sources cause one burst per
year, i.e. the number of source events produced in the interval
$\Delta t$ is relative to 1 year observation time.

In order to estimate the event rates for non-persistent sources it
is crucial to know their duty cycle. Some of these sources have a
periodic bursting activity: Circinus X-1 has a period of 16.59
days \cite{Preston83}. This means therefore that we expect about
1.5 events per year. Other transient sources show a stochastic
bursting activity. For such cases it is difficult to give an
estimate of the expected event rate. For example, during 1994 GRO
J1655-40 had three radio flares, each lasting 6 days
\cite{Hjellming95}; during the same year GRS 1915+105 emitted 4
bursts \cite{Rodriguez99}. A recent study \cite{Nipoti05} has
shown that microquasars GRS 1915+105, Cygnus X-3 and Scorpius X-1
are in flaring mode 21, 10 and 3 percent of the time,
respectively. The possibility to integrate over more then one
burst could therefore help to detect neutrinos from microquasars.

The search for neutrino events in coincidence with microquasar
radio outbursts could be a tool to reject atmospheric background,
restricting the analysis period to the flare duration $\Delta t$.
Such an analysis technique, already used by AMANDA
\cite{Amanda05}, can improve the detector sensitivity to neutrinos
from transient sources. Referring to the bursts considered in Tab.
\ref{tab:N_mu_simulated} and integrating over the time interval
$\Delta t$ of the bursts, we expect an average background of about
$10^{-3}$ events (muons) per burst. Summing on all the bursting
sources in Tab. \ref{tab:N_mu_simulated}, we count $\sim0.04$
background events, which requires about 5 source events for a
5$\sigma$ level detection with a 70\% probability \cite{IceCube}.
Tab. \ref{tab:N_mu_simulated} shows that we expect $3.4\div9.0$
events in the case of a burst from each of the bursting
microquasars. Therefore, a cumulative analysis could provide a
possible detection of microquasar neutrinos.

For the microquasar LS 5039 we considered the flux predicted by
Aharonian et al. \cite{Aharonian05}. The expected number of
selected events is given in Tab. \ref{tab:N_mu_aharonian}; the
comparison with the atmospheric background shows that an evidence
could be expected in a few years of data taking.

In Tab. \ref{tab:N_mu_bednarek}  we quote the expected number of events
from microquasar Cygnus X-3, according to the Bednarek model \cite{Bednarek05}.
Since the model predicts a neutrino flux with a spectral index close to 2
and since the main event contribution is in the energy range 1 - 100 TeV,
we used the same event selection parameters quoted in Tab. \ref{tab:sensitivity}.
In the case of $\kappa=2.5$ and $\gamma_A^{max}=10^7$ the signal is lower than the background.
In the other cases, the number of events exceeds the expected background. The signal could be
therefore detected in a few years of data taking, especially in the case of
$\kappa=2$ and $\gamma_A^{max}=10^7$ with 2.2 events per year.

\begin{table}[h]
\renewcommand{\arraystretch}{1.2}
\begin{center}
\caption{{\bf Expected number of neutrino induced muons from the Levinson and Waxman microquasar model:}
N$_\mu^{m}$ is the number of selected muons from each
microquasar expected from the theoretical neutrino energy flux
$f_\nu^{th}$ quoted by Distefano et al. \cite{Distefano02}, during the time interval $\Delta t$.
We also report the expected number of atmospheric background events $b$ surviving the
event selection and expected in 1 year of data taking.}
\label{tab:N_mu_simulated}
~\\*[0.2cm]
\begin{tabular}{lcc|cc}
\hline
     {\bf Source name}       & {\bf ${\bf \Delta t}$ (days)} & {\bf ${\bf f_\nu^{th}}$ (erg/cm$^2$ s)} & {\bf ${\bf N_\mu^{m}}$} &  {\bf ${\bf b}$}  \\
\hline
\hline
{\it Steady Sources} \\
\hline
  LS 5039                 & 365            & 1.69$\cdot10^{-12}$                        &   0.1     		&   0.1      	\\
  Scorpius X-1            & 365            & 6.48$\cdot10^{-12}$                        &   0.2     		&   0.1      	\\
  SS433                   & 365            & 1.72$\cdot10^{-9~}$                        &   76.0        	&   0.1        	\\
  GX 339-4                & 365            & 1.26$\cdot10^{-9~}$                        &   68.0        	&   0.1        	\\
  Cygnus X-1              & 365            & 1.88$\cdot10^{-11}$                        &   0.5        	 	&   0.1    	\\
\hline
{\it Bursting Sources} \\
\hline
  XTE J1748-288           & 20             & 3.07$\cdot10^{-10}$                        &   0.8     		&   0.3        	\\
  Cygnus X-3              & 3              & 4.02$\cdot10^{-9~}$                        &   0.8     		&   0.1    	\\
  GRO J1655-40            & 6              & 7.37$\cdot10^{-10}$                        &   0.6     		&   0.1        	\\
  GRS 1915+105            & 6              & 2.10$\cdot10^{-10}$                        &   0.1     		&   $<0.1$   	\\
  Circinus X-1            & 4              & 1.22$\cdot10^{-10}$                        &   0.1     		&   0.1        	\\
  XTE J1550-564           & 5              & 2.00$\cdot10^{-11}$                        &   $<0.1$      	&   $<0.1$  	\\
  V4641 Sgr               & 0.3            & 2.25$\cdot10^{-10}\div3.25\cdot10^{-8}$    &   $<0.1\div$1.4   	&   0.1       	\\
  GS 1354-64              & 2.8            & 1.88$\cdot10^{-11}$                        &   $<0.1$      	&   0.1     	\\
  GRO J0422+32            & 1$\div$20      & 2.51$\cdot10^{-10}$                        &   $<0.1\div$0.4   	&   0.1      	\\
  XTE J1118+480           & 30$\div$150    & 5.02$\cdot10^{-10}$                        &   1.0$\div$4.8    	&   0.2      	\\
\hline
\end{tabular}
\\*[1.cm]
\end{center}
\end{table}

\begin{table}
\renewcommand{\arraystretch}{1.2}
\begin{center}
\caption{{\bf Expected number of neutrino induced muons from microquasar LS 5039 (Aharonian et al. model \cite{Aharonian05}):}
we give the number of selected events N$_\mu^{m}$ from LS 5039 compared to the atmospheric background events $b$
surviving the event selection, during an interval time of 1 year.}
\label{tab:N_mu_aharonian}
~\\*[0.2cm]
\begin{tabular}{cc|cc}
\hline
\hline
\multicolumn{4}{c}{{\bf LS 5039}}\\
\hline
     {\bf ${\bf \Gamma}$} & {\bf ${\bf \varepsilon_\nu^{max}}$ (TeV)} &    {\bf ${\bf N_\mu^{m}}$} &  {\bf ${\bf b}$}\\
\hline
\hline
  1.5            & ~10        & 1.7  & 0.2      \\
  1.5            & 100        & 4.9  & 0.1      \\
  2.0            & ~10        & 1.0  & 0.3      \\
  2.0            & 100        & 2.6  & 0.1      \\
\hline
\end{tabular}
\\*[1.cm]
\end{center}
\end{table}

\begin{table}
\renewcommand{\arraystretch}{1.2}
\begin{center}
\caption{{\bf Expected number of neutrino induced events from microquasar Cygnus X-3 (Bednarek model \cite{Bednarek05}):}
N$_\mu^{m}$ is the number of selected muons from Cygnus X-3
during an interval time of 1 year.
We also report the number of atmospheric background muons $b$ surviving the event selection during the same interval time.}
\label{tab:N_mu_bednarek}
~\\*[0.2cm]
\begin{tabular}{c|ccc}
\hline
\hline
  {\bf Cygnus X-3} &     {\bf REG. II}   &  {\bf REG. III} &   \\
\hline
     {\bf ${\bf ~~~\kappa,~\gamma_A^{max}}$}              & {\bf N${\bf _\mu^{m}}$}   & {\bf N${\bf _\mu^{m}}$}  &  {\bf ${\bf b}$} \\
\hline
  $2.0,~10^6$             & 0.4        &  0.4       	&  0.1     	\\
  $2.0,~10^7$             & 1.1        &  1.1       	&  0.1     	\\
  $2.5,~10^7$             & $<0.1$     &  0.1   	&  0.1      	\\
mono-energetic            & $<0.1$     &  0.7   	&  0.1      	\\
\hline
\end{tabular}
\\*[1.cm]
\end{center}
\end{table}

\section{Conclusions}

The possibility to detect TeV neutrinos
from Galactic microquasars with the proposed NEMO-km$^3$ underwater
\v{C}erenkov neutrino telescope has been investigated.

A Monte Carlo was carried out to simulate the expected
neutrino-induced muon fluxes produced by microquasars and by
atmospheric neutrinos. The expected atmospheric muon background
was also simulated. Muon tracks were propagated inside the
detector using the ANTARES simulation tools, taking into account
the water optical parameters measured in the Capo Passero site.
Muon tracks were then reconstructed and the number of neutrino
induced muons expected from microquasars was determined, according
to different theoretical neutrino flux predictions quoted in the
literature. These results were compared with the number of
expected background events from atmospheric neutrinos and muons.
We applied an event selection in order to reject the atmospheric
background. The applied event selection is a combination of
different criteria, chosen optimizing the detector sensitivity to
neutrino fluxes coming from each microquasar. Eventually we
calculated the expected number of events surviving the selection
referring to 1 year of data taking.

We made a comparison of our sensitivity to microquasar/point-like
sources with the analysis made by IceCube
\cite{IceCube,barcellona}. For a 3 years exposure the
sensitivities for a $\varepsilon_\nu^{-2}$ energy dependence are
$\sim10^{-9}$ and $\sim 2\cdot 10^{-9}$ GeV/cm$^2$ s for NEMO and
IceCube, respectively.

Our results show that, assuming reasonable scenarios for TeV
neutrino production, the proposed NEMO telescope could identify
microquasars in a few years of data taking,
%with a discovery potential for at least few cases above the 5$\sigma$ level,
with the possibility of a 5$\sigma$ level detection;
otherwise it would strongly constrain the neutrino production models and the source
parameters.

\section*{Acknowledgments}

We thank the ANTARES Collaboration for providing the
detector simulation and track reconstruction codes, extensively used in this work.
We thank also F.A. Aharonian, W. Bednarek and G.E. Romero
for discussions and comments.

\end{document}